\documentclass[12pt]{article}
\usepackage{a4wide}
\usepackage{epsfig}
\usepackage{amsmath}

\newlength{\absize}
\catcode`@=11
\def\citer{\@ifnextchar [{\@tempswatrue\@citexr}{\@tempswafalse\@citexr[]}}

\def\@citexr[#1]#2{\if@filesw\immediate
  \write\@auxout{\string\citation{#2}}\fi
  \def\@citea{}\@cite{\@for\@citeb:=#2\do
    {\@citea\def\@citea{--\penalty\@m}\@ifundefined
       {b@\@citeb}{{\bf ?}\@warning
       {Citation `\@citeb' on page \thepage \space undefined}}%
\hbox{\csname b@\@citeb\endcsname}}}{#1}}
\catcode`@=12

\begin{document}
  \thispagestyle{empty}
  \pagestyle{empty}
  \renewcommand{\thefootnote}{\fnsymbol{footnote}}
\newpage\normalsize
    \pagestyle{plain}
    \setlength{\baselineskip}{4ex}\par
    \setcounter{footnote}{0}
    \renewcommand{\thefootnote}{\arabic{footnote}}
\newcommand{\preprint}[1]{%
  \begin{flushright}
    \setlength{\baselineskip}{3ex} #1
  \end{flushright}}
\renewcommand{\title}[1]{%
  \begin{center}
    \LARGE #1
  \end{center}\par}
\renewcommand{\author}[1]{%
  \vspace{2ex}
  {\Large
   \begin{center}
     \setlength{\baselineskip}{3ex} #1 \par
   \end{center}}}
\renewcommand{\thanks}[1]{\footnote{#1}}
\begin{flushright}
%July 9, 2005
\end{flushright}
\vskip 0.5cm

\begin{center}
{\large \bf Searching for Effects of Spatial Noncommutativity via
Chern-Simons' Processes$\;^{\ast}$}
\end{center}
\vspace{1cm}
\begin{center}
Jian-Zu Zhang%$\;^{\ast}$
\end{center}
%-----------------------------------
%   Address
%-----------------------------------
\vspace{1cm}
\begin{center}
Institute for Theoretical Physics, East China University of
Science and Technology, Box 316, Shanghai 200237, P. R. China
\end{center}
\vspace{1cm}
%%%%%%%%%%%%%%%%%%%%%%%%%%%%%%%%%%%%%%%%%%%%%%%%%%%%%%%%%%%%%%

\begin{abstract}
The possibility of testing spatial noncommutativity in the case of
both position-position and momentum-momentum noncommuting via a
Chern-Simons' process is explored. A Chern-Simons process can be
realized by an interaction of a charged particle in special
crossed electric and magnetic fields, in which the Chern-Simons
term leads to non-trivial dynamics in the limit of vanishing
kinetic energy. Spatial noncommutativity leads to the spectrum of
the orbital angular momentum possessing fractional values.
Furthermore, in both limits of vanishing kinetic energy and
subsequent vanishing magnetic field, the Chern-Simons term leads
to this system having non-trivial dynamics again, and the dominant
value of the lowest orbital angular momentum being $\hbar/4$,
which is a clear signal of spatial noncommutativity. An
experimental verification of this prediction by a
Stern-Gerlach-type experiment is suggested.
\end{abstract}

\begin{flushleft}
$^{\ast}$ Extended version of hep-ph/0508042
\end{flushleft}
%\begin{flushleft}
%$^{\ast}$ E-mail: jzzhang@ecust.edu.cn
%\end{flushleft}
%%%%%%
%\begin{flushleft}
%$^{a)}$ Extended version of hep-th/0508164
%\end{flushleft}
\clearpage
%%%%%%%%%%%%%%%%%%%%%%%%%%%%%%%%%%%%%%%%%%%%%%%%%%%%%%%%%%%%%%%%%
{\bf 1. Introduction}

\vspace{0.4cm}

Studies of low energy effective theory of superstrings show that
space is noncommutative \citer{CDS,DN}. Spatial non-commutativity
is apparent near the Planck scale. Its modifications to ordinary
quantum theory are extremely small. We ask whether one can find
some low energy detectable relics of physics at the Planck scale
by current experiments. Such a possibility is inferred from the
incomplete decoupling between effects at high and low energy
scales. For the purpose of clarifying phenomenological low energy
effects, quantum mechanics in noncommutative space (NCQM) is
available. If NCQM is a realistic physics, all the low energy
quantum phenomena should be reformulated in it. In literature,
NCQM have been studied in detail \citer{CST,JZZ04a}; many
interesting topics, from the Aharonov-Bohm effect to the quantum
Hall effect have been considered \citer{CDPST00,LD}. Recent
investigations of the deformed Heisenberg-Weyl algebra (the NCQM
algebra) in noncommutative space explore some new features of
effects of spatial noncommutativity \cite{JZZ04a}. The possibility
of testing spatial noncommutativity via Rydberg atoms is explored.
But there are two problems in the suggested experiment of Rydberg
atoms: (1) The special arrangement of the electric field required
in the experiment is difficult to realized in laboratories; (2)
The measurement depends on a extremely high characteristic
frequency which may be difficult to reach by current experiments.

In this paper we show a possibility of testing spatial
noncommutativity via a Chern-Simons process. Chern-Simons'
processes \citer{DJT89,DJT90} exhibit interesting properties in
physics. In laboratories a Chern-Simons process can be realized by
an interaction of a charged particle in special crossed electric
and magnetic fields, in which the experimental situation is
different from one in the experiment of Rydberg atoms. Properties
of the Chern-Simons process at the level of NCQM are investigated.
Spatial noncommutativity leads to the spectrum of the orbital
angular momentum possessing a fractional zero-point angular
momentum. In the limit of vanishing kinetic energy the
Chern-Simons term leads to this system having non-trivial
dynamics. For the case of both position-position and
momentum-momentum noncommuting in a further limit of the
subsequent diminishing magnetic field this system possesses
non-trivial dynamics again, and the dominant value of the lowest
orbital angular momentum in the process is $\hbar/4$. This result
is a clear signal of spatial noncommutativity, and can be verified
by a Stern-Gerlach-type experiment, in which two difficulties in
the experiment of Rydberg atoms are resolved.

In Ref.~\citer{GJPP} other electromagnetic effects of spatial
noncommutativity was explored.

\vspace{0.4cm}

{\bf 2. The Deformed Heisenberg-Weyl algebra}

\vspace{0.4cm}

In the following we review the background of the deformed
Heisenberg-Weyl Algebra. In order to develop the NCQM formulation
we need to specify the phase space and the Hilbert space on which
operators act. The Hilbert space can consistently be taken to be
exactly the same as the Hilbert space of the corresponding
commutative system \citer{CST}.

As for the phase space we consider both position-position
noncommutativity (position-time noncommutativity is not
considered) and momentum-momentum noncommutativity. There are
different types of noncommutative theories, for example, see a
review paper \cite{DN}.

In the case of both position-position and momentum-momentum
noncommuting the consistent deformed Heisenberg-Weyl algebra
\cite{JZZ04a} is:
\begin{equation}
\label{Eq:xp}%1e
[\hat x_{I},\hat x_{J}]=i\xi^2\theta_{IJ}, \qquad [\hat x_{I},\hat
p_{J}]=i\hbar\delta_{IJ}, \qquad [\hat p_{I},\hat
p_{J}]=i\xi^2\eta_{IJ},\;(I,J=1,2,3)
\end{equation}
where $\theta_{IJ}$ and $\eta_{IJ}$ are the antisymmetric constant
parameters, independent of the position and momentum. We define
$\theta_{IJ}=\epsilon_{IJK}\theta_K$ (Henceforth the summation
convention is used), where $\epsilon_{IJK}$ is a three-dimensional
antisymmetric unit tensor. We put $\theta_3=\theta$ and the rest
of the $\theta$-components to zero (which can be done by a
rotation of coordinates), then we have
$\theta_{ij}=\epsilon_{ij}\theta$ $(i,j=1,2)$, where
$\epsilon_{ij3}$ is rewritten as a two-dimensional antisymmetric
unit tensor $\epsilon_{ij}$, $\epsilon_{12}=-\epsilon_{21}=1,$
$\epsilon_{11}=\epsilon_{22}=0$. Similarly, we have
$\eta_{ij}=\epsilon_{ij}\eta$. In Eqs.~(\ref{Eq:xp}) the scaling
factor $\xi$ is
%%%%%%
\begin{equation}
\label{Eq:xi-1}%2e
\xi=(1+\frac{1}{4\hbar^2}\theta\eta)^{-1/2}.
\end{equation}
%%%%%%
It plays a role for guaranteeing consistent representations of
$(\hat x_i, \hat p_j)$ in terms of the undeformed canonical
variables $(x_i, p_j )$ (See Eqs.~(\ref{Eq:hat-x-x})).

In noncommutative space questions about whether the concept of
identical particles being still meaningful and whether
Bose-Einstein statistics being still maintained should be
answered. Bose - Einstein statistics can be investigated at two
levels: the level of quantum field theory and the level of quantum
mechanics. On the fundamental level of quantum field theory the
annihilation and creation operators appear in the expansion of the
(free) field operator $\Psi(\hat{x})=\int d^3k
a_k(t)\Phi_k(\hat{x})+h.c.$. The consistent multi-particle
interpretation requires the usual (anti)commutation relations
among $a_k$ and $a^\dagger_k$. Introduction of the Moyal type
deformation of coordinates may yield a deformation of the algebra
between the creation and annihilation operators \cite{BMPV}.
Whether the deformed Heisenberg - Weyl algebra is consistent with
Bose - Einstein statistics is an open issue at the level of
quantum field theory.

In this paper our study is restricted in the context of
non-relativistic quantum mechanics. Following the standard
procedure in the ordinary quantum mechanics in commutative space
we construct the deformed annihilation-creation operators $(\hat
a_i$, $\hat a_i^\dagger)$ which are related to the deformed
canonical variables $(\hat x_i, \hat p_i)$. In order to maintain
the physical meaning of $\hat a_i$ and $\hat a_i^\dagger$ the
relations among $(\hat a_i, \hat a_i^\dagger)$ and $(\hat x_i,
\hat p_i)$ should keep the same formulation as the ones in
commutative space. For a system with mass $\mu$ and frequency
$\omega=\omega_p/2$ (Here the reason of introducing $\omega_p/2$
is that in the Hamiltonian (\ref{Eq:H-2}) the potential energy
takes the same form as one of harmonic oscillator) the $\hat a_i$
reads
\begin{equation}
\label{Eq:aa+1}%3e
\hat a_i=\sqrt{\frac{\mu\omega_p}{4\hbar}}\left (\hat x_i
+i\frac{2}{\mu\omega_p}\hat p_i\right).
\end{equation}
%%%%%%
From Eq.~(\ref{Eq:aa+1}) and the deformed Heisenberg-Weyl algebra
(\ref{Eq:xp}) we obtain the commutation relation between the
operators $\hat a_{i}$ and $\hat a_{j}$: $\left[\hat a_i,\hat
a_j\right]=i\xi^{2}\mu\omega_p\epsilon_{ij}\left (\theta-4
\eta/\mu^{2}\omega^{2}_p\right)/4\hbar$.
%$\equiv D_{ij}$.
When the
state vector space of identical bosons is constructed by
generalizing one-particle quantum mechanics, in order to maintain
Bose-Einstein statistics at the deformed level described by $\hat
a_i$ the basic assumption is that operators $\hat a_i$ and $\hat
a_j$ should be commuting. This requirement leads to a consistency
condition
\begin{equation}
\label{Eq:cc}%4e
\eta=\frac{1}{4}\mu^2\omega_p^2 \theta.
\end{equation}
%%%%%%
%It guarantees consistency of the deformed Heisenberg-Weyl algebra
%with Bose-Einstein statistics, and
which puts constraint between the parameters $\eta$ to $\theta$.
The commutation relations of $\hat a_i$ and $\hat a_j^\dagger$ are
%%%%%%
\begin{equation}
\label{Eq:[a,a+]1}%5e
[\hat a_i,\hat
a_j^\dagger]=\delta_{ij}+i\frac{1}{2\hbar}\xi^2\mu\omega_p
\theta\epsilon_{ij},\; [\hat a_i,\hat a_j]=0.
\end{equation}
%%%%%%
%\begin{equation}
%\label{Eq:[a,a+]1}%?e
%[\hat a_1,\hat a_1^\dagger]=[\hat a_2,\hat a_2^\dagger]=1, [\hat
%a_1,\hat a_2]=0;\quad [\hat a_1,\hat a_2^\dagger]
%=i\xi^2\mu\omega_p \theta/2\hbar.
%\end{equation}
Here, the three equations $[\hat a_1,\hat a_1^\dagger]=[\hat
a_2,\hat a_2^\dagger]=1,\; [\hat a_1,\hat a_2]=0$ are the same
boson algebra as the one in commutative space. The equation
%%%%%%
\begin{equation}
\label{Eq:[a,a+]2}%6e
[\hat a_1,\hat a_2^\dagger] =i\frac{1}{2\hbar}\xi^2\mu\omega_p
\theta
\end{equation}
%%%%%%
is a new type. Different from the case in commutative space, it
correlates different degrees of freedom to each other, so it is
called the correlated boson commutation relation. It encodes
effects of spatial noncommutativity at the deformed level
described by $(\hat a_i, \hat a_j^\dagger)$, and plays essential
roles in dynamics \cite{JZZ04a}.

It is worth noting that Eq.~(\ref{Eq:[a,a+]2}) is consistent with
{\it all} principles of quantum mechanics and Bose-Einstein
statistics.

If momentum-momentum were commuting, $\eta= 0$, we could not
obtain $[\hat a_i,\hat a_j]=0$. It is clear that in order to
maintain Bose-Einstein statistics for identical bosons at the
deformed level we should consider both position-position
noncommutativity and momentum-momentum noncommutativity. In this
paper momentum-momentum noncommutativity means the {\it intrinsic}
noncommutativity. It differs from the momentum-momentum
noncommutativity in an external magnetic field; In that case the
corresponding noncommutative parameter is determined by the
external magnetic field. Here both parameters $\eta$ and $\theta$
should be extremely small, which is guaranteed by the consistency
condition (\ref{Eq:cc}).

The deformed Heisenberg-Weyl algebra (\ref{Eq:xp}) has different
realizations by undeformed variables $(x_i,p_i)$ \cite{NP}. We
consider the following consistent ansatz of a linear
representation of the deformed variables $(\hat x_i, \hat p_j)$ by
the undeformed variables $(x_i, p_j)$:
\begin{equation}
\label{Eq:hat-x-x}%7e
\hat x_{i}=\xi(x_{i}-\frac{1}{2\hbar}\theta\epsilon_{ij}p_{j}
 ), \quad \hat
p_{i}=\xi(p_{i}+\frac{1}{2\hbar}\eta\epsilon_{ij}x_{j}).
\end{equation}
where $x_{i}$ and $p_{i}$ satisfy the undeformed Heisenberg-Weyl
algebra $[x_{i},x_{j}]=[p_{i},p_{j}]=0,
[x_{i},p_{j}]=i\hbar\delta_{ij}.$ It is worth noting that the
scaling factor $\xi$ is necessary for guaranteeing that the
Heisenberg commutation relation $[\hat x_{i},\hat
p_{j}]=i\hbar\delta_{ij}$ is maintained by Eq.~(\ref{Eq:hat-x-x}).

The last paper in Ref.~\cite{JZZ04a} clarified that though the
deformed
%phase space variables
$\hat x_{i}$ and $\hat p_{j}$ are related to the
undeformed $x_{i}$ and $p_{j}$ by the linear transformation
(\ref{Eq:hat-x-x}), the deformed Heisenberg-Weyl algebra is
related to the undeformed one by a similarity transformation with
a non-orthogonal real matrix and a unitary similarity
transformation which transforms two algebras to each other does
not exist, thus two algebras are not unitarily equivalent.

\vspace{0.4cm}

{\bf 3. Chern-Simons' Interactions}

\vspace{0.4cm}

%%%%%%
Physical systems confined to a space-time of less than four
dimensions show a variety of interesting properties. There are
well-known examples, such as the quantum Hall effect, high $T_c$
superconductivity, cosmic string in planar gravity, etc. In many
of these cases the Chern-Simons interaction \citer{DJT89,DJT90},
which exists in 2+1 dimensions and is associated with the
topologically massive gauge fields, plays a crucial role. In
laboratories a Chern-Simons' process can be realized by an
interaction of a charged particle in special crossed electric and
magnetic fields, an example is a Penning trap \citer{BG,Kost}, in
which an analog of the Chern-Simons term reads
%%%%%%
%\begin{equation}
%\label{Eq:V}%e
$$\epsilon_{ij}\hat x_i\hat p_j.$$
%\end{equation}
%%%%%%
This term leads to non-trivial dynamics in the limit of vanishing
kinetic energy, and in turn a testable effect of spatial
noncommutativity.
%%%%%%

The objects trapped in a Penning trap are charged particles or
ions. The trapping mechanism combines an electrostatic quadrupole
potential
\begin{equation}
\label{Eq:V}%8e
\hat \phi=\frac{V_0}{2d^2}(-\frac{1}{2}\hat x_i^2+\hat x_3^2),
(i=1, 2)
\end{equation}
and a uniform magnetic field ${\bf B}$ aligned along the $z$ axis.
%%%%%%
The vector potential $\hat A_i$ corresponding to the uniform
magnetic field ${B}$ reads
\begin{equation}
\label{Eq:A}%9e
\hat A_i=\frac{1}{2}\epsilon_{ij}B\hat x_j.
\end{equation}
The parameters $V_0(>0)$ and $d$ are the characteristic voltage
and length. The particle oscillates harmonically with an axial
frequency $\omega_z=(qV_0/\mu d^2)^{1/2}$ (charge $q>0$) along the
axial direction (the $z$-axis), and in the (1, 2) - plane,
executes a superposition of a fast circular cyclotron motion of a
cyclotron frequency $\omega_c=qB/\mu c$ with a small radius, and a
slow circular magnetron drift motion of a magnetron frequency
$\omega_m\equiv \omega_z^2/2\omega_c$ in a large orbit. Typically
the quadrupole potential superimposed upon the magnetic field is a
relatively weak addition in the sense that the hierarchy of
frequencies is
%%%%%%
\begin{equation}
\label{Eq:hierarchy}%10e
\omega_m<<\omega_z<<\omega_c.
\end{equation}
%%%%%%
The Hamiltonian $\hat H$ of this system can be decomposed into a
two-dimensional Hamiltonian $\hat H_2$ and a one-dimensional
harmonic Hamiltonian $\hat H_z$:
\begin{equation}
\label{Eq:H}%11e
\hat H=\frac{1}{2\mu}(\hat p_i-\frac{q}{c}\hat A_i)^2+q\hat
\phi=\hat H_2+\hat H_z,
\end{equation}
%%%%%%
\begin{equation}
\label{Eq:H-z}%12e
 \hat H_z= \frac{1}{2\mu}\hat p_3^2
+\frac{1}{2}\mu\omega_z^2\hat x_3^2,
\end{equation}
and $\hat H_2$ is \cite{BG,Dehm}
\begin{equation}
\label{Eq:H-2}%13e
\hat H_2= \frac{1}{2\mu}\hat p_i^2
+\frac{1}{8}\mu\omega_p^2\hat x_i^2
-\frac{1}{2}\omega_c\epsilon_{ij}\hat x_i\hat p_j,
\end{equation}
where $\mu$ is the particle mass, $\omega_p\equiv
\omega_c(1-4\omega_m/\omega_c)^{1/2}$. If NCQM is a realistic
physics, low energy quantum phenomena should be reformulated in
this framework. In the above the noncommutative Hamiltonian
(\ref{Eq:H-2}) is obtained by reformulating the corresponding
commutative one $H_2 = p_i^2/2\mu + \mu\omega_p^2 x_i^2/8 -
\omega_c\epsilon_{ij}x_i p_j/2$ in commutative space in terms of
the deformed canonical variables $\hat x_i$ and $\hat p_i$.

In Eq.~(\ref{Eq:H-2}) the term $\omega_c\epsilon_{ij}\hat x_i\hat
p_j/2$ plays an interesting role of realizing analogs of the
Chern-Simons theory \citer{DJT89,DJT90}.

In order to explore the new features of such a system our
attention is focused on the investigation of $\hat H_2$ and the
$z$ component of the orbital angular momentum. There are different
ways to define the deformed angular momentum in noncommutative
space.
%%%%%%%%%%%%%%%%%%%%%%%%%%%%%%%%%%%%%%%%%%%%%%%%%%%%%%%%%   note-2
%\footnote{\;

(i) As a generator of rotations at the deformed level
the deformed angular momentum $\hat J^{\prime}_z$ should transform
$\hat x_i$ and $\hat p_j$ as two dimensional vectors \cite{NP}:
%%%%%%
$$[\hat J^{\prime}_z,\hat x_i]=i\epsilon_{ij}\hat
x_j,\;[\hat J^{\prime}_z,\hat p_i]=i\epsilon_{ij}\hat p_j.$$
%%%%%%
Comparing to the case in commutative space, the deformed angular
momentum $\hat J^{\prime}_z$ acquires $\theta-$ and
$\eta-$dependent scalar terms $\hat x_i \hat x_i$ and $\hat p_i
\hat p_i$,
%%%%%%
\begin{equation}
\label{Eq:Ja}%14e
\hat J^{\prime}_z=\frac{\hbar^2}{\hbar^2-\xi^4\theta\eta}
\left(\epsilon_{ij}\hat x_i\hat p_j+\frac{\xi^2\eta}{2\hbar}\hat
x_i\hat x_i+\frac{\xi^2\theta}{2\hbar}\hat p_i\hat p_i\right).
\end{equation}
%%%%%%

(ii) The quantum mechanical system described by the deformed
Hamiltonian Eq.~(\ref{Eq:H-2}), or equivalently
Eq.~(\ref{Eq:CS-H1}), possesses a full rotational symmetry in (1,
2) - plane. The generator of those rotations is given as
\begin{equation}
\label{Eq:Jb}%15e
J_z=\epsilon_{ij}x_ip_j,
\end{equation}
i.e., all quantities $\hat{x_i}, \hat{p_i}, x_i, p_i$ transforms
as two dimensional vectors.

(iii) The third point of view is as follows: If NCQM is a
realistic physics, all deformed observables (the deformed
Hamiltonian, the deformed angular momentum, etc.) in
noncommutative space can be obtained by reformulating the
corresponding undeformed ones in commutative space in terms of
deformed canonical variables. Thus the deformed angular momentum
$\hat J_z$, like the deformed Hamiltonian (\ref{Eq:H-2}), keeps
the same representation as the undeformed one $J_z$, but is
reformulated in terms of $\hat x_i$ and $\hat p_i$, i.e., the
Chern-Simons term
%%%%%%
\begin{equation}
\label{Eq:J}%16e
\hat J_z=\epsilon_{ij}\hat x_i\hat p_j.
\end{equation}
%%%%%%

Our starting point is at the deformed level. Because of the scalar
terms in $\hat J^{\prime}_z$ have nothing to do with the angular
momentum, so in this paper we prefer to take Eq.~(\ref{Eq:J}) as
the definition of $\hat J_z$. Eq.~(\ref{Eq:[a,a+]2}) modifies the
commutation relations between $\hat J_z$ and $\hat x_i,$ $\hat
p_i$. From the NCQM algebra (\ref{Eq:xp}) we obtain
\begin{equation*}%
[\hat J_z,\hat x_i]=i\epsilon_{ij}\hat x_j+i\xi^{2}\theta \;\hat
p_i, \quad
%%%%%%
 [\hat J_z,\hat
p_i]=i\epsilon_{ij}\hat p_j-i\xi^{2} \eta\;\hat x_i.\nonumber
\end{equation*}
Comparing with the commutative case, the above commutation
relations acquires $\theta-$ and $\eta-$dependent terms which
represent effects in noncommutative space. From the above
commutation relations we conclude that $\hat J_z$ plays
approximately the role of the generator of rotations at the
deformed level.

All quantities $\hat J_z$, $\hat J^{\prime}_z$, $J_z$ and $\hat
H_2$ commute each other, thus have common eigenstates.
%}
%%%%%%%%%%%%%%%%%%%%%%%%%%%%%%%%%%%%%%%%%%%%%%%%%%%%%%%%%%%%%%%
For example, from Eqs.~(\ref{Eq:xp}), (\ref{Eq:cc}),
(\ref{Eq:H-2}) and (\ref{Eq:J}) it follows that
\begin{equation}
\label{Eq:H-J}%17e
[\hat J_z,\hat H_2]=0.
\end{equation}
Here cancellations between $\theta-$ and $\eta-$ dependent terms,
provided by the consistency condition (\ref{Eq:cc}), is crucial
for obtaining Eq.~(\ref{Eq:H-J}).

The  $\theta-$ and $\eta-$ dependent terms of the Chern-Simons
term $\hat J_z$ have no direct relation to the angular momentum,
see Eq.~(\ref{Eq:J1}). The essential point is whether the
interaction with magnetic field (the Stern-Gerlach part of the
apparatus) is mediated by $J_z$ or $\hat{J}_z$. The definition of
a generator of rotations elucidates that the interaction with
magnetic field is mediated by $J_z$. It is worth noting that in
both limits of vanishing kinetic energy and subsequent vanishing
magnetic field $\hat{J}_z$ is proportional to $J_z$, see
Eqs.~(\ref{Eq:J02})-(\ref{Eq:J03}). This means that in the
particular limit the eigenvalue of the Chern-Simons term $\hat
J_z$ should appear in the spectrum of angular momentum.

$\hat H_2$ and $\hat J_z$ constitute a complete set of observables
of the two-dimensional sub-system. Using Eqs.~(\ref{Eq:hat-x-x})
the Hamiltonian $\hat H_2$ is represented by undeformed variables
$x_i$ and $p_i$ as
\begin{eqnarray}
\label{Eq:CS-H1}%18e
\hat H_2=\frac{1}{2M}(p_i+\frac{1}{2}G\epsilon_{ij} x_j)^2
-\frac{1}{2}K x_i^2 =\frac{1}{2M} p_i^2+\frac{1}{2M}G\epsilon_{ij}
p_i x_j+\frac{1}{8}M\Omega_p^2 x_i^2,
\end{eqnarray}
where the effective parameters %$M, \Omega_c, \Omega_p$
$M, G, \Omega_p$ and $K$ are defined as
%%%%%%
\begin{eqnarray}
\label{Eq:M-K}%19e
1/M&\equiv& \xi^2\left(b_1^2/\mu-qV_0
\theta^{\;2}/8d^2\hbar^2\right),\;
%%%%%%
%\Omega_c
G/M\equiv \xi^2\left(2b_1 b_2/\mu-qV_0
\theta/2d^2\hbar\right),\nonumber\\
%%%%%%
M\Omega_p^2&\equiv& \xi^2\left(4b_2^2/\mu-2qV_0/d^2 \right),\;
%%%%%%
%K\equiv M(\Omega_c^2-\Omega_p^2)/4,
K\equiv (G^2/M-M\Omega_p^2)/4,
\end{eqnarray}
and $b_1=1+qB\theta/4c\hbar,\;b_2=qB/2c+\eta/2\hbar.$ The
parameter $K$ consists of the difference of two terms. It is worth
noting that the dominant value of $K$ is
$qV_0/2d^2=\mu\omega_z^2/2$, which is positive.

%%%%%%
Similarly, from Eq.~(\ref{Eq:J}) and Eqs.~(\ref{Eq:hat-x-x}) the
Chern-Simons term $\hat J_z$ is rewritten as
\begin{equation}
\label{Eq:J1}%20e
\hat J_z=\epsilon_{ij}x_ip_j-\frac{1}{2\hbar}\xi^{2}\left(\theta
p_i p_i +\eta x_i
x_i\right)=J_z-\frac{1}{2\hbar}\xi^{2}\left(\theta p_i p_i +\eta
x_i x_i\right).
\end{equation}
%%%%%%

The deformed Heisenberg-Weyl algebra and the undeformed one are,
respectively, the foundations of noncommutative and commutative
quantum theories. Because of the unitary un-equivalency between
two algebras it is expected that the spectrum of deformed
observables (the Hamiltonian $\hat H_2$, the angular momentum
$\hat J_z$, etc.) may be different from the spectrum of the
corresponding undeformed ones ($H_2$, $J_z$, etc.).

\vspace{0.4cm}

{\bf 4. Dynamics in the limiting case of vanishing kinetic energy}

\vspace{0.4cm}

In the following we are interested in the system  (\ref{Eq:CS-H1})
for the limiting case of vanishing kinetic energy. In this limit
the Hamiltonian (\ref{Eq:CS-H1}) has non-trivial dynamics, and
there are constraints which should be carefully considered
\cite{JZZ04a,Baxt,JZZ96}. For this purpose it is more convenient
to work in the Lagrangian formalism. The limit of vanishing
kinetic energy in the Hamiltonian formalism identifies with the
limit of the mass $M\to 0$ in the Lagrangian formalism.
%\cite{foot1}.
%%%%%%%%%%%%%%%%%%%%%%%%%%%%%%%%%%%%%%%%%%%%%%%%%%%%%%%%%%%%%%%
%\bibitem{foot1}
%\footnote {}
In Eq.~(\ref{Eq:CS-H1}) in the limit of vanishing kinetic energy,
$\frac{1}{2M}\left( p_i+\frac{1}{2}
%M\Omega_c
G\epsilon_{ij} x_j\right)^2=\frac{1}{2}M \dot{x_i} \dot{x_i}\to
0,$ the Hamiltonian $\hat H_2$ reduces to
%%%%%%
\begin{equation}
\label{Eq:H0}%21e
H_0=-\frac{1}{2}K x_i x_i.
\end{equation}
%%%%%%
The Lagrangian corresponding to the
Hamiltonian (\ref{Eq:CS-H1}) is
%%%%%%
\begin{equation}
\label{Eq:L}%22e
L=\frac{1}{2}M\dot{x_i}\dot{x_i} -\frac{1}{2}
%M\Omega_c
G\epsilon_{ij}\dot{x_i}x_j+\frac{1}{2}K x_i x_i.
\end{equation}
%%%%%%
In the limit of $M\to 0$ this Lagrangian reduces to
%%%%%%
\begin{equation}
\label{Eq:L0}%23e
L_0=-\frac{1}{2}
%M\Omega_c
G\epsilon_{ij}\dot{x_i}x_j +\frac{1}{2}K x_i x_i.
\end{equation}
%%%%%%
From $L_0$ the corresponding canonical momentum is
$p_{0i}=\partial L_0/\partial \dot{x_i}=-\frac{1}{2}
%M\Omega_c
G\epsilon_{ji}x_j,$ and the corresponding Hamiltonian is
$H_0^{\prime}=p_{0i}\dot{x_i}-L_0=-\frac{1}{2}K x_i x_i=H_0.$ Thus
we identify the two limiting processes.
%%%%%%%%%%%%%%%%%%%%%%%%%%%%%%%%%%%%%%%%%%%%%%%%%%%%%%%%%%%%%%%%%
Here the point is that when the potential is velocity dependent
the limit of vanishing kinetic energy in the Hamiltonian does not
corresponds to the limit of vanishing velocity in the Lagrangian.
If the velocity approached zero in the Lagrangian, there would be
no way to define the corresponding canonical momentum, thus there
would be no dynamics.

The massless limit have been studied by Dunne, Jackiw and
Trugenberger \cite{DJT90}.

The first equation of (\ref{Eq:CS-H1}) shows that in the limit
$M\to 0$ there are constraints
%%%%%%%%%%%%%%%%%%%%%%%%%%%%%%%%%%%%%%%%%%%%%%%%%%%%%%%%%%%%%%
%\footnote{\;In this example the symplectic method \cite{FJ} leads
%to the same results as the Dirac method for constrained
%quantization, and the representation of the symplectic method is
%much streamlined.}
%%%%%%%%%%%%%%%%%%%%%%%%%%%%%%%%%%%%%%%%%%%%%%%%%%%%%%%%%%%%%%
\begin{equation}
\label{Eq:Ci}%24e
C_i=p_i+\frac{1}{2}
%M\Omega_c
G\epsilon_{ij} x_j=0,
\end{equation}
which should be carefully treated. In this example the Faddeev-
Jackiw's symplectic method \cite{FJ} leads to the same results as
the Dirac method for constrained quantization, and the
representation of the symplectic method is much streamlined. In
the following we adopt the Dirac method \cite{MZ}. The Poisson
brackets
of the constraints %(\ref{Eq:Ci})
are $\{C_i,C_j\}_P=
%M\Omega_c
G\epsilon_{ij}\ne 0,$ so that the corresponding Dirac brackets of
the canonical variables $x_i, p_j$ can be determined,
\begin{equation}
\label{Eq:D,x-p}%25e
\{x_i,p_j\}_D=\frac{1}{2}\delta_{ij},\;
\{x_1,x_2\}_D=-\frac{1}
%{M\Omega_c}
G,\;\{p_1,p_2\}_D= -\frac{1}{4}
%M\Omega_c
G.
\end{equation}
The Dirac brackets of $C_i$  with any variables $x_i$  and $p_j$
are
zero so that the constraints %(\ref{Eq:Ci})
are strong conditions, and can be used to eliminate the dependent
variables. If we select $x_1$ and $p_1$ as the independent
variables, from the constraints %(\ref{Eq:Ci})
we obtain $x_2=-2p_1/
%M\Omega_c
G,\;p_2=
%M\Omega_c
Gx_1/2.$ We introduce new canonical variables $q=\sqrt{2}x_1$  and
$p=\sqrt{2}p_1$ which satisfy the Heisenberg quantization
condition $[q,p]=i\hbar$, and define the effective mass
$\mu^{\ast}$ and the effective frequency $\omega^{\ast}$ as
%%%%%%
\begin{equation}
\label{Eq:omega-ast}%26e
\mu^{\ast}\equiv \frac{
%M^2\Omega_c^2
G^2}{2K},\;
%%%%%%
\omega^{\ast} \equiv \frac{K}{
%M\Omega_c
G},
\end{equation}
%%%%%%
then the Hamiltonian $\hat H_2$ reduces to
\begin{equation}
\label{Eq:H0a}%27e
H_0=-\left(\frac{1}{2\mu^{\ast}}p^2+
\frac{1}{2}\mu^{\ast}\omega^{\ast2 }q^2\right).
\end{equation}

We define an annihilation operator
%%%%%%
\begin{equation}
\label{Eq:A}%28e
A= \sqrt{\frac{\mu^{\ast}\omega^{\ast}}{2\hbar}}\;q
+i\sqrt{\frac{1}{2\hbar\mu^{\ast}\omega^{\ast}}}\;p,
\end{equation}
%%%%%%
The annihilation and creation operators $A$ and $A^\dagger$
satisfies $[A,A^\dagger]=1$, and the eigenvalues of the number
operator $N=A^\dagger A$ is $n^{\prime}=0, 1, 2, \cdots$.

The Hamiltonian $H_0$ is rewritten as
%%%%%%
\begin{equation}
\label{Eq:H0b}%29e
H_0=-\hbar\omega^{\ast}\left(A^\dagger A+\frac{1}{2}\right).
\end{equation}
%%%%%%
Similarly, the angular momentum $J_z$ and the Chern-Simons term
$\hat J_z$ reduce, respectively, to the following $J_z^{\prime}$
and $\hat J_z^{\prime}$
%%%%%%
\begin{equation}
\label{Eq:Jz}%30e
J_z^{\prime}=\hbar\left(A^\dagger A+\frac{1}{2}\right),\;
%\end{equation}
%%%%%%
%%%%%%
%\begin{equation}
%\label{Eq:Jz}%?e
\hat J_z^{\prime}=\hbar\mathcal{J}^{\ast}\left(A^\dagger
A+\frac{1}{2}\right)=\mathcal{J}^{\ast}J_z^{\prime},
\end{equation}
%%%%%%
where
%%%%%%
\begin{equation}
\label{Eq:Jz1}%31e
\mathcal{J}^{\ast}=1-\xi^2\left(\frac{
%M\Omega_c
G\theta}{4\hbar} +\frac{\eta}{
%M\Omega_c
G\hbar}\right).
\end{equation}
%%%%%%
The eigenvalues of $H_0$ and $\hat J_z$ are, respectively,
%%%%%%
%\begin{eqnarray}
\begin{equation}
\label{Eq:En}%32e
E_n^{\ast}=-\hbar\omega^{\ast}\left(n^{\prime}+\frac{1}{2}\right),
\end{equation}
%\nonumber\\
\begin{equation}
\label{Eq:Jzn}%33e
\mathcal{J}_n^{\ast}=\hbar\mathcal{J}^{\ast}\left(n^{\prime}
+\frac{1}{2}\right),
\end{equation}
%\end{eqnarray}
%%%%%%
The eigenvalue of $H_0$ is negative, thus unbound. This motion is
unstable. It is worth noting that the dominant value of
$\omega^{\ast}$ is the magnetron frequency $\omega_m$, i.e. in the
limit of vanishing kinetic energy the surviving motion is
magnetron-like, which is more than adequately metastable \cite
{BG,Dehm}.

The $\theta-$ and $\eta-$ dependent terms of $\mathcal{J}^{\ast}$
take fractional values. Thus the Chern - Simons term $\hat J_z$
possesses fractional eigenvalues and fractional intervals.

Using the consistency condition (\ref{Eq:cc}) we rewrite the
$\mathcal{J}^{\ast}$ in Eq.~(\ref{Eq:Jz1}) as
%%%%%%
%\begin{equation}
%\label{Eq:Jz2}%?e
$\mathcal{J^{\ast}}=1+O(\theta)$.
%\end{equation}
%%%%%%
From Eq.~(\ref{Eq:Jzn}) it follows that the zero-point value
$\mathcal{J}_0^{\ast}$ reads
%%%%%%
\begin{equation}
\label{Eq:Jz0}%34e
\mathcal{J}_0^{\ast}=\frac{1}{2}\hbar+O(\theta).
\end{equation}
%%%%%%

For the case of both position-position and momentum-momentum
noncommuting we can consider a further limiting process. After the
sign of $V_0$ is changed, the definition of $\Omega_p$ shows that
the limit of magnetic field $B\to 0$ is meaningful, and the
survived system also has non-trivial dynamics. In this limit the
frequency $\omega_p$ reduces to
%%%%%%
$\tilde \omega_p=\sqrt{2}\omega_z$,
%%%%%%
the consistency condition (\ref{Eq:cc}) becomes a reduced
consistency condition
%%%%%%
\begin{equation}
\label{Eq:cc1}%35e
\eta=\frac{1}{2}\mu^2\omega_z^2\theta,
\end{equation}
%%%%%%
and the scaling parameter $\xi$ in Eq.~(\ref{Eq:xi-1}) reduces to
%%%%%%
\begin{equation}
\label{Eq:xi-2}%36e
\tilde \xi
=\left(1+\frac{1}{8\hbar^2}\mu^2\omega_z^2\theta^2\right)^{-1/2}
=1+O(\theta^2).
\end{equation}
%%%%%%
The effective parameters $M,
%\Omega_c
G, \Omega_p$ and $K$ reduce, respectively, to the following
effective parameters $\tilde M, \tilde
%\Omega_c
G, \tilde \Omega_p$ and $\tilde K$, which are defined by
%%%%%%
\begin{eqnarray}
\label{Eq:M-K1}%37e
\tilde M &\equiv& \left[\tilde \xi^2 \left(\frac{1}{\mu} +
\frac{1}{8\hbar^2}\mu\omega_z^2\theta^2\right)\right]^{-1}= \mu,
\nonumber\\
%%%%%%
%\Omega_c
\frac{\tilde G}{\tilde M}&\equiv& \tilde
\xi^2\left(\frac{\eta}{\mu\hbar}+
\frac{1}{2\hbar}\mu\omega_z^2\theta\right)=
\frac{1}{\hbar}\mu\omega_z^2\theta+O(\theta^3),
\nonumber\\
%%%%%%
\tilde \Omega_p^2&\equiv& \tilde \xi^2\left(
\frac{\eta^2}{\mu^2\hbar^2}+2\omega_z^2\right)=2\omega_z^2+O(\theta^2),
\nonumber\\
%%%%%%
\tilde K&\equiv&
%\frac{1}{4}\tilde M\left(\tilde \Omega_c^2-\tilde
%\Omega_p^2\right)
\frac{1}{4}\left(\frac{\tilde G^2}{\tilde M}-\tilde M\tilde
\Omega_p^2\right) =-\frac{1}{2}\mu\omega_z^2+O(\theta^2).
\end{eqnarray}
%%%%%%
In this limit $H_0$ and $\hat J_z^{\prime}$ reduce, respectively,
to the following $\tilde H_0$ and $\tilde J_z$:
\begin{equation}
\label{Eq:H02}%38e
\tilde H_0=-\frac{1}{2}\tilde K x_i^2=\hbar\tilde
\omega\left(\tilde A^\dagger \tilde A+\frac{1}{2}\right),
\end{equation}
%\nonumber\\
%%%%%%%%%%%%%%%%%%%%%%%%%%%%%%%%%%%%%%%%%%%%%%%%%%%%%%%%%%%%%%%%
\begin{equation}
\label{Eq:J02}%39e
\tilde J_z=\hbar\mathcal{\tilde J}\left(\tilde A^\dagger \tilde
A+\frac{1}{2}\right)=\mathcal{\tilde J}J_z^{\prime}.
\end{equation}
%%%%%%
In this limit the angular momentum $J_z^{\prime}$ is not changed,
but can be rewritten as
\begin{equation}
\label{Eq:Jz02}%40e
J_z^{\prime}=\hbar\left(\tilde A^\dagger \tilde
A+\frac{1}{2}\right).
\end{equation}
%%%%%%
In Eq.~(\ref{Eq:J02})
\begin{equation}
\label{Eq:J03}%41e
\mathcal{\tilde J}=1-\tilde \xi^2\left(\frac{1}{4\hbar}
%\tilde
%M\tilde \Omega_c
\tilde G\theta +\frac{\eta}{
%\tilde M\tilde \Omega_c
\tilde G\hbar}\right).
\end{equation}
In the above the annihilation operator is defined as
%%%%%%
\begin{equation}
\label{Eq:A1}%42e
\tilde A=\sqrt{\frac{\tilde \mu\tilde \omega}{2\hbar}}\;q
+i\sqrt{\frac{1}{2\hbar\tilde \mu\tilde \omega}}\;p,
\end{equation}
%%%%%%
and the effective mass $\tilde \mu$ and the effective frequency
$\tilde \omega$ are
\begin{equation}
\label{Eq:tilde-omega}%43e
\tilde \mu\equiv -\frac{\tilde
%M^2\tilde \Omega_c^2
G^2}{2\tilde K}\left(>0\right),\;
%\tilde \omega^2 \equiv
%\left(\frac{\tilde K}{
%\tilde M\tilde \Omega_c
%G}\right)^2.
\tilde \omega \equiv \frac{\tilde K}{\tilde G}.
\end{equation}
%%%%%%%
The annihilation and creation operators $\tilde A$ and $\tilde
A^\dagger$ satisfies $[\tilde A,\tilde A^\dagger]=1$, and the
eigenvalues of the number operator $\tilde N=\tilde A^\dagger
\tilde A$ is $n=0, 1, 2, \cdots$.
%%%%%%

Eqs.~(\ref{Eq:H02})-(\ref{Eq:J03}) show that $\tilde H_0$, $\tilde
J_z$ and $J_z^{\prime}$ commute each other, thus have common
eigenstates.
%%%%%%
The eigenvalues of $\tilde H_0$, $\tilde J_z$ and $J_z^{\prime}$
are, respectively,
%%%%%%
%\begin{eqnarray}
%\label{Eq:En-Jn}%?e
\begin{equation}
\label{Eq:En}%44e
\tilde E_n=\hbar\tilde\omega\left(n+\frac{1}{2}\right),\;
%\nonumber\\
\end{equation}
%%%%%%
\begin{equation}
\label{Eq:Jn}%45e
\mathcal{\tilde J}_n=\hbar\mathcal{\tilde
J}\left(n+\frac{1}{2}\right),\;
%%%%%%
\mathcal{J^{\prime}}_n=\hbar\left(n+\frac{1}{2}\right).
\end{equation}
%\end{eqnarray}
%%%%%%
It is worth noting that in both limits of vanishing kinetic energy
and subsequent vanishing magnetic field we have
%%%%%%
$\mathcal{\tilde J}_n=\mathcal{\tilde J}
\mathcal{J^{\prime}}_n$.
%%%%%%

Now we estimate the dominant value of the constant
$\mathcal{\tilde J}$. A dominant value of an observable means its
$\theta-$ and $\eta-$ independent term. Generally, the dominant
value is just the value in commutative space. In some special case
the consistency condition (\ref{Eq:cc}) or the reduced consistency
condition (\ref{Eq:cc1}) may provides a cancellation between
$\theta$ and $\eta$ in some term of an observable. This leads to
that the dominant value is different from one in commutative
space.
%%%%%%

In the third term of $\mathcal{\tilde J}$ in Eq.~(\ref{Eq:J03}),
unlike the term $\eta/
%M\Omega_c
G\hbar=O(\eta)$ of $\mathcal{J}^{\ast}$ in Eq.~(\ref{Eq:Jz1}),
the reduced consistency condition (\ref{Eq:cc1}) provides a fine
cancellation between $\theta$ and $\eta$. Using
Eqs.~(\ref{Eq:cc1}) - (\ref{Eq:M-K1}), this term reads
%%%%%%
$\eta/
%\tilde M\tilde \Omega_c
\tilde G\hbar=1/2$,
%%%%%%
which leads to
%%%%%%
\begin{equation}
\label{Eq:J3}%46e
\mathcal{\tilde J}=\frac{1}{2}+O(\theta^2).
\end{equation}
%%%%%%
From Eqs.~(\ref{Eq:Jn}) and (\ref{Eq:J3}) it follows that the
zero-point value $\mathcal{\tilde J}_0$ is
%%%%%%
\begin{equation}
\label{Eq:J0}%47e
\mathcal{\tilde J}_0=\frac{1}{4}\hbar+O(\theta^2),
\end{equation}
%%%%%%
and the interval $\Delta \mathcal{\tilde J}_n$ of the Chern-Simons
term reads
%%%%%%
\begin{equation}
\label{Eq:J0}%48e
\Delta \mathcal{\tilde J}_n=\frac{1}{2}\hbar+O(\theta^2).
\end{equation}
%%%%%%
The dominant values of the zero-point value and the interval of
the Chern-Simons term are, respectively, $\hbar/4$ and $\hbar/2$,
which are different from the values in commutative space. These
unusual results explore the essential new feature of spatial
noncommutativity.

\vspace{0.4cm}

{\bf 5. Testing Spatial Noncommutativity via a Penning Trap}

\vspace{0.4cm}

The dominant value $\hbar/4$ of the lowest Chern - Simons term in
a Penning trap can be measured by a Stern-Gerlach-type experiment.
The experiment consists of two parts: the trapping region and the
Stern-Gerlach experimental region. The trapping region serves as a
source of the particles for the Stern-Gerlach experimental region.
After establishing the trap, the experiment includes three steps.

(i) Taking the limit of vanishing kinetic energy. In an
appropriate laser trapping field the speed of atoms can be slowed
to the extent that the kinetic energy term may be removed \cite
{SST}. In the limit of vanishing kinetic energy the situations of
the cyclotron motion,  the harmonic axial oscillation and the
magnetron-like motion in a Penning trap are different \cite
{BG,Dehm}.
%%%%%%%%%%%%%%%%%%%%%%%%%%%%%%%%%%%%%%%%%%%%%%%%%%%%%%%%%
%\footnote{\; }
The energy in the cyclotron motion is almost exclusively kinetic
energy. The energy in the harmonic axial oscillation alternates
between kinetic and potential energy. Reducing the kinetic energy
in either of these motions reduces their amplitude. In contrast to
these two motions, the energy in the magnetron-like motion is
almost exclusively potential energy. Thus in the limit of
vanishing kinetic energy the harmonic axial oscillation and the
cyclotron motion disappear, only the magnetron-like motion
survives. Any process that removes energy from the magnetron-like
motion increases the magnetron radius until the particle strikes
the ring electrode and is lost from the trap. The magnetron-like
motion is unstable. Fortunately, its damping time is on the order
of years \cite {BG}, so that it is more than adequately
metastable. In this limit, the survived magnetron-like motion
slowly drifts in a large orbit in the (1, 2) - plane. At the
quantum level, in the limit of vanishing kinetic energy the mode
with the frequency $\omega^{\ast}$ survives. As we noted before,
the dominant value of $\omega^{\ast}$ is the magnetron frequency
$\omega_m$, i.e. the surviving mode is magnetron-like.

(ii) Changing the sign of the voltage $V_0$ and subsequently
diminishing the magnetic field $B$ to zero. The voltage $V_0$ is
weak enough so that when the magnetic field $B$ approaches zero
the trapped particles can escape along the tangent direction of
the circle from the trapping region and are injected into the
Stern-Gerlach experimental region.

%%%%%%
(iii) Measuring the $z$-component of the lowest Chern - Simons
term in the Stern-Gerlach experimental region. As noticing before,
the commutation relations between $\hat J_z$ and $\hat x_i,$ $\hat
p_i$ show that $\hat J_z$ plays approximately the role of the
generator of rotations at the deformed level. Eqs.~(\ref{Eq:H02})
- (\ref{Eq:J03}) and (\ref{Eq:Jn}) elucidate that the lowest
dominant value $\hbar/4$ of the Chern - Simons term $\tilde J_z$
can be read out from spectrum of the angular momentum which are
measured from the deflection of the beam in the Stern-Gerlach
experimental region.
%%%%%%%%%%%%%%%%%%%%%%%%%%%%%%%%%%%%%%%%%%%%%%%%%%%%%%%%  note-3

\vspace{0.4cm}

{\bf 6. Discussions}

\vspace{0.4cm}

As is well known,
%according the uncertainty principle,
a direct measurement of the magnetism and the gyromagnetic ratio
for free electrons are impossible.
%In order to make observations
%of such properties possible, electrons should be bound with large
%mass, so that the disturbance of the experimental instrument is
%small.
Thus in the above suggested experiment, the trapped object are
chosen as ions.

When ions are injected into the Stern-Gerlach experimental region,
in order to avoid a disturbance of the Lorentz motion in the
inhomogeneous magnetic field, they should first go through a
region of revival and are restored to neutral atoms. We should
choose ions of the first class atoms in periodic table of the
elements.
% for example, ions of the Cs atoms.
An advantage of choosing such ions is that in the ordinary case of
commutative space the revived atoms are in the $S$-state.

Now we estimate the possibility of changing the state of revived
atoms by effects of spatial noncommutativity. The effective
frequency $\tilde \omega$ in Eq.~(\ref{Eq:tilde-omega}) depends on
noncommutative parameters. There are different bounds on the
parameter $\theta$ set by experiments. The existing experiments on
the Lorentz symmetry violation placed strong bounds on $\theta$
\cite{CHKLO}: $\theta/(\hbar c)^2\le (10 \;TeV)^{-2}$;
Measurements of the Lamb shift \cite{CST} give a weaker bound;
Clock-comparison experiments \cite{MPR} claim a stronger bound.
The magnitudes of $\theta$ and $\eta$ are surely extremely small.
From Eq.~(\ref{Eq:tilde-omega}) it follows that the dominate value
of the frequency $\tilde \omega$ reads $\tilde \omega = |\tilde
K|/
%\tilde M\tilde \Omega_c
\tilde G\approx \hbar/(2\mu\theta)$. If we take $\mu c^2=2 \,GeV$
and $\theta/(\hbar c)^2\le (10^4 \;GeV)^{-2}$ we obtain $\tilde
\omega \ge 10^{32} Hz$. Eq.~(\ref{Eq:En}) shows that the
corresponding energy interval $\Delta\tilde E_n$ is extremely
large. Thus revived atoms can not transit to higher exciting
states, they are definitely preserved in the ground state.

%%%%%%
The result obtained in this paper is different from results
obtained in literature. All effects of spatial noncommutativity
explored in literature depend on extremely small noncommutative
parameters $\theta$ and/or $\eta$, thus can not be tested in the
foreseeable future. Because of a direct proportionality between
$\theta$ and $\eta$ provided by the reduced consistency condition
(\ref{Eq:cc1}), in Eq.~(\ref{Eq:J03}) there is a fine cancellation
between $\theta$ and $\eta$. This leads to a $\theta-$ and $\eta-$
independent effect of spatial noncommutativity which can be tested
by current experiments.

In both limits of vanishing kinetic energy and subsequent
diminishing magnetic field for the case of only position-position
noncommuting dynamics of the Penning trap is trivial; But for the
case of both position-position and momentum-momentum noncommuting
its dynamics is non-trivial, and the dominant value $\hbar/4$ of
the lowest Chern - Simons term in the Penning trap is different
from the value in commutative space. The above suggested
experiment can distinguish the case of both position-position and
momentum-momentum noncommuting from the case of only
position-position noncommuting.

\vspace{0.4cm}

{\bf Acknowledgements}

\vspace{0.4cm}

The author would like to thank Jean-Patrick Connerade, bai-Wen Li,
Ming-Sheng Zhan, Ting-Yun Shi, Si-Hong Gu and Qing-Yu Cai for
helpful discussions. This work has been partly supported by the
National Natural Science Foundation of China under the grant
number 10575037 and by the Shanghai Education Development
Foundation.

%\clearpage

\end{document}